\documentclass[twocolumn, english,superscriptaddress,
aps, showpacs, showkeys, amsmath, amssymb, floatfix, nofootinbib]{revtex4-1}

\pdfoutput=1
\usepackage[pdftex]{graphicx}
\usepackage[pdftex]{color}
\usepackage[colorlinks,bookmarks]{hyperref}
\definecolor{linkblue}{rgb}{0,0,0.8}
\definecolor{linkgreen}{rgb}{0,0.5,0}
\hypersetup{linkcolor=linkblue, citecolor=linkgreen, urlcolor=linkblue}

\usepackage[T1]{fontenc}
\usepackage[utf8]{inputenc}
\usepackage{lmodern}
\usepackage{babel}
\usepackage{bm}
\usepackage{esint}
\usepackage{verbatim}
\usepackage[us]{datetime}
\usepackage{booktabs}
\usepackage{enumerate}
\usepackage{graphics}
\usepackage{float}
\usepackage{amsmath}
\usepackage{amsfonts}
\usepackage{amssymb}

\def\d{{\rm d}}
\def\L{\mathcal{L}}
\def\E{\mathcal{E}}

\definecolor{vale}{rgb}{0,0.5, 1.}
\definecolor{david}{rgb}{0.74, 0, 0.58}

\begin{document}

\title{Local determination of the Hubble constant and the deceleration parameter}

\author{David Camarena}
\affiliation{PPGCosmo, Universidade Federal do Espírito Santo, 29075-910, Vitória, ES, Brazil}

\author{Valerio Marra}
\affiliation{Núcleo Cosmo-ufes \& Departamento de Física, Universidade Federal do Espírito Santo, 29075-910, Vitória, ES, Brazil}

\begin{abstract}
The determination of the Hubble constant $H_0$ from %the well-understood physics of 
the Cosmic Microwave Background by the Planck Collaboration~\cite{Aghanim:2018eyx}  is in tension at $4.2\sigma$ with respect to the local determination of $H_0$ by the SH0ES collaboration~\cite{Reid:2019tiq}. %, which uses local Supernovae calibrated via the cosmic distance ladder.
Here, we improve upon the local determination, which fixes the deceleration parameter  to the standard $\Lambda$CDM model value of $q_0=-0.55$, that is, uses information from observations beyond the local universe.
First, we derive the effective calibration prior on the absolute magnitude $M_B$ of Supernovae Ia, which can be used in cosmological analyses in order to avoid the double counting of low-redshift supernovae.
We find $M_B = -19.2334 \pm 0.0404$~mag.
Then, we use the above $M_B$ prior in order to obtain a determination of the local $H_0$ which only uses local observations and only assumes the cosmological principle, that is, large-scale homogeneity and isotropy.
This is achieved by adopting an uninformative flat prior for $q_0$ in the cosmographic expansion of the luminosity distance.
We use the latest Pantheon sample and find $H_0= 75.35 \pm 1.68 \text{ km s}^{-1} {\rm Mpc}^{-1}$, which features a 2.2\% uncertainty, close to the 1.9\% error obtained by the SH0ES Collaboration.
Our determination is at the higher tension of $4.5\sigma$ with the latest results from the Planck Collaboration that assume the $\Lambda$CDM  model.
Furthermore, we also constrain the deceleration parameter to $q_0= -1.08 \pm 0.29$, which  
disagrees with Planck at the $1.9\sigma$ level.
These estimations only use supernovae in the redshift range $0.023\le z\le 0.15$.
\end{abstract}

\keywords{Hubble constant, Cosmology}

\maketitle

%%%%%%%%%%%%%%%%%%%%%%%%%%%%%%%%%%%%
%%%%%%%%%%%%%%%%%%%%%%%%%%%%%%%%%%%%
\section{Introduction}

The $\Lambda$CDM model -- the standard model of cosmology -- has been extremely successful.
Assuming only General Relativity and well-understood linear perturbations about a homogeneous and isotropic background model, it accounts, with just 6 parameters, for basically all cosmological observations on a vast range of scales in space and time.
Its key ingredients are the cosmological constant, a constant of nature, and the dark matter,
a yet-undetected particle which is predicted by, e.g., supersymmetric extensions of the standard model of particle physics.
Although the theoretical basis for both the cosmological constant and dark matter may be rightfully questioned, the standard model has been keeping pragmatically its throne, thanks to its performance and simplicity.

However, since the first release of the Cosmic Microwave Background (CMB) observations by the Planck Collaboration in 2013 \cite{Ade:2013zuv}, the determination of the Hubble  constant $H_0$ based on the standard model of cosmology started to be in tension with the model-independent determination via calibrated local Supernovae Ia (SN) by the SH0ES Collaboration in 2011~\cite{Riess:2011yx}.
The initial tension of $2.4\sigma$ worsened over the past 6 years. On one side, systematics were better understood and CMB data accumulated~\cite{Spergel:2013rxa,Aghanim:2016yuo}. On the other side, the sample of local supernovae increased and the anchors used to calibrate them considerably improved~\cite{2019Natur.567..200P,Riess:2019cxk,Reid:2019tiq}.
The present situation is that the two determinations of the Hubble   constant --  one by the Planck Collaboration in 2018~\cite{Aghanim:2018eyx} and the other by the SH0ES Collaboration in late 2019~\cite{Reid:2019tiq} -- are now in tension at the considerable $4.2\sigma$ level.
See~\cite{Freedman:2017yms} for a historical overview % of the last 20 years of $H_0$ determinations
and~\cite{Verde:2019ivm} for the present status.

Much work has been done trying to understand the implications of this tension: it is indeed, by far, the most severe problem the $\Lambda$CDM model is facing.
On one hand, the effect of the local structure -- the so-called cosmic variance on $H_0$ -- has been thoroughly studied (see \cite{Camarena:2018nbr} and references therein), as well as possible re-assessments of the error budget \cite{Cardona:2016ems,Zhang:2017aqn,Feeney:2017sgx}.
On the other hand, physics beyond the standard model has been investigated, hoping that this tension could shine a light on possible alternatives to the highly tuned cosmological constant and the yet-undetected dark matter \cite[see][for example]{Vattis:2019efj,Poulin:2018cxd,Guo:2018ans,Colgain:2018wgk,DiValentino:2016hlg}.
However, at the moment, it is not clear which kind of physics beyond $\Lambda$CDM could solve this crisis~\cite{Knox:2019rjx}.

Here, we wish to improve upon the local determination by the SH0ES Collaboration, which adopts a Dirac delta prior on the deceleration parameter, centered at the standard $\Lambda$CDM model value of  $q_0=-0.55$.
In \cite{Feeney:2017sgx} it was shown that using the broad (truncated) Gaussian prior  $q_0=-0.5\pm 1$ it is indeed possible to obtain a competitive constraint on the Hubble constant.
In what follows, first we derive the calibration prior on the absolute magnitude $M_B$ of Supernovae Ia that was effectively used in the $H_0$ determination by the SH0ES Collaboration.
Then, using the $M_B$ prior and an uninformative flat prior on $q_0$, we will obtain a competitive determination of the  $H_0$.
Indeed, our determination has an uncertainty comparable with the one by SH0ES and it is at the even higher tension of $4.5\sigma$ with the latest results from the Planck Collaboration that assume the $\Lambda$CDM  model.
We stress that our determination only uses local observations and only assumes the FLRW metric, that is, the cosmological principle, according to which the universe is homogeneous and isotropic at large scales.

This paper is organized as follows.
After introducing low-redshift cosmography in Section~\ref{sec:cg}, we will obtain the effective calibration prior in Section~\ref{sec:cp} and the determination of the Hubble constant in Section~\ref{sec:h0} using the Pantheon dataset.
We will discuss our results in Section~\ref{sec:disc} and conclude in Section~\ref{sec:conc}.
Finally, Appendix~\ref{sec:dm} presents details regarding the derivation of the effective calibration prior, Appendix~\ref{sec:sc} reports the results relative to the Supercal supernova dataset, and Appendix~\ref{sec:R19} reports the results relative to the early-2019 determination of $H_0$ by~\cite{Riess:2019cxk}.

%%%%%%%%%%%%%%%%%%%%%%%%%%%%%%%%%%%%
%%%%%%%%%%%%%%%%%%%%%%%%%%%%%%%%%%%%
\section{Cosmography} \label{sec:cg}

The apparent magnitude $m_B^t$ of a supernova at redshift $z$ is given by
\begin{equation} \label{mB1}
m_B^t(z) = 5\log_{10}\left[\frac{d_L(z)}{1 \text{Mpc}} \right] + 25 + M_B \,, 
\end{equation}
where $d_L(z)$ is the luminosity distance and $M_B$ the absolute magnitude. Using a cosmographic approach within an FLRW metric -- which only assumes large-scale homogeneity and isotropy -- one has:
\begin{align} \label{dL1}
d_L(z) &=  \frac{c z}{H_0} \left [ 1 + \frac{(1-q_0)z}{2}  + O(z^2) \right] \,,
\end{align}
where the Hubble   constant and the deceleration parameter are defined, respectively, according to:
\begin{align}
H_0=\left. \frac{\dot a(t)}{a(t)} \right|_{t_0} \,,
\qquad
q_0=\left. \frac{- \ddot a(t)}{H(t)^2 a(t)} \right|_{t_0}  \,.
\end{align}
Cosmography is a model-independent approach in the sense that it does not assume a specific model as it is based on the Taylor expansion of the scale factor. However, this does not mean that its parameters do not contain cosmological information.
For example, the deceleration parameter is connected to the parameters of $w$CDM cosmologies according to:
\begin{align}
q_0 &= \frac{\Omega_{\rm m 0}}{2} + \frac{1+3 w}{2} \Omega_{\rm de 0}
= \frac{1 + 3 w\,  \Omega_{\rm de 0}}{2} \nonumber \\
&= \frac{1 - 3 \, \Omega_{\Lambda 0}}{2} \simeq -0.55 \,,  \label{q0} 
\end{align}
where the last equality in first line assumes spatial flatness ($\Omega_{\rm m 0}+\Omega_{\rm de 0}=1$),
the first equality in the second line assumes the $\Lambda$CDM model ($w=-1$), and the last one the concordance value of $\Omega_{\Lambda 0} \simeq 0.7$.
%The identification of equation \eqref{q0} is clearly model dependent.

\begin{figure}
\includegraphics[width= \columnwidth]{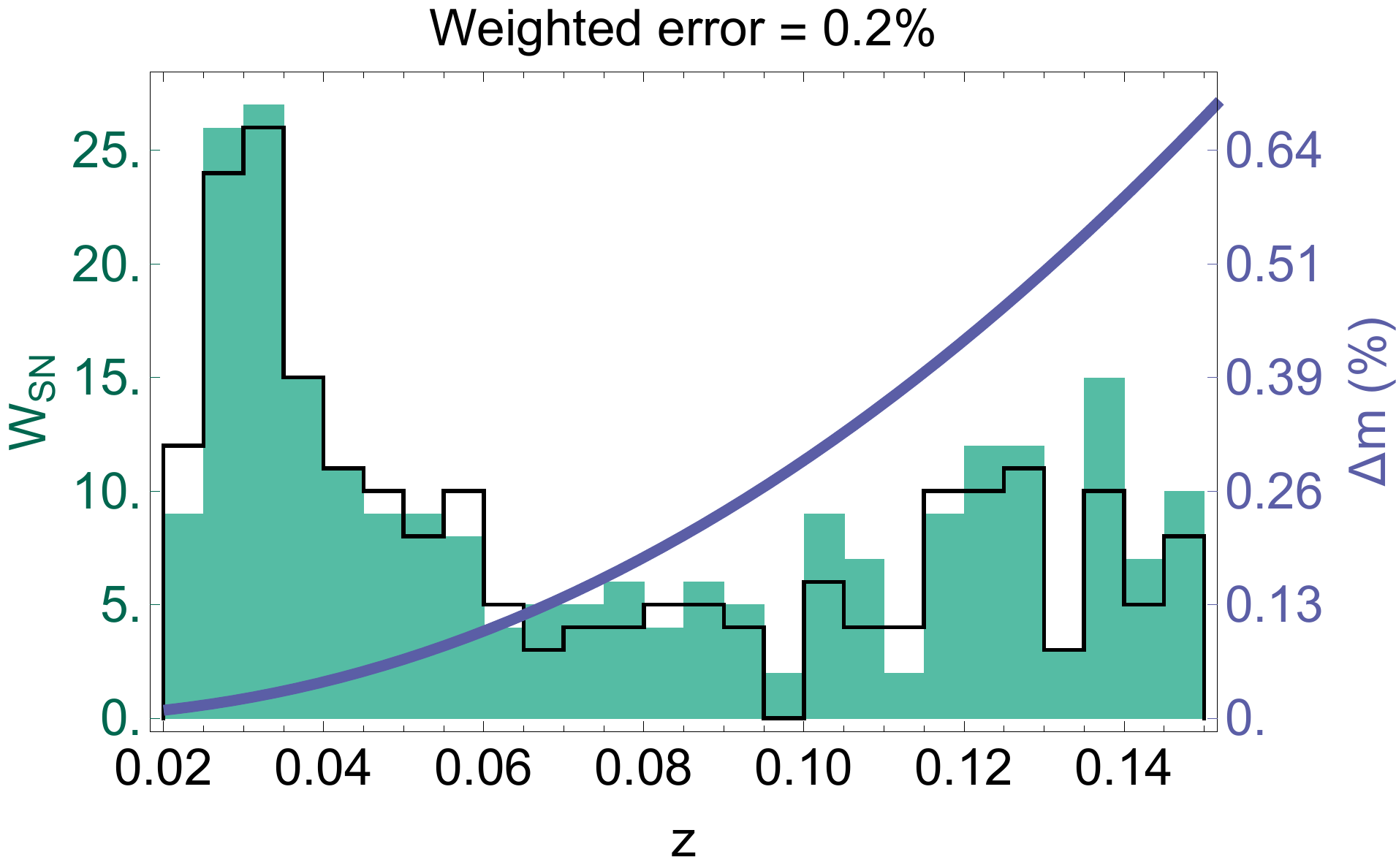}
\caption{Shown with a solid line is the percentage difference between the distance modulus with and without the second-order correction in the expansion of the luminosity distance of equation \eqref{dL1}. The difference is always smaller than 0.7\%. The histogram shows the distribution $W_{SN}$ of the supernovae in the range $0.023\le z\le 0.15$ that are used for local determinations of $H_0$.
The filled green histogram corresponds to the Pantheon sample (237 supernovae)~\cite{Scolnic:2017caz} while the empty black one to the Supercal sample (217 suepernovae)~\cite{Scolnic:2015eyc}.
Using these distributions the weighted error from neglecting the second-order correction is only $0.2\%$.}
\label{z2}
\end{figure}

Combining now equations (\ref{mB1}-\ref{dL1}) one has:
\begin{align}
\nonumber m_B^t & =  5 \log_{10} \left[ 1 + \frac{(1-q_0)z}{2} + O(z^2)  \right] \\
& + 5 \log_{10}{\frac{c z H_0^{-1}}{\text{1 Mpc}}}  + 25 + M_B \,. \label{mB2}
\end{align}
We are neglecting the second-order correction that contains the jerk parameter $j_0=\left.\frac{\dddot a}{H^3 a}\right|_{t_0}$ because we will consider only low-redshift supernovae.
Indeed, the latest local determinations of $H_0$ use supernovae in the range $0.023\le z\le 0.15$, where the minimum redshift is large enough in order to reduce the impact of cosmic variance \cite{Marra:2013rba,Camarena:2018nbr} and the maximum redshift is small enough in order to reduce the impact of cosmology in the determination of $H_0$.
We have to bear in mind that the cosmographic expansion could be problematic if extended to high redshifts ($z>1$). Indeed, it fails to converge \cite{Cattoen:2007sk} and also its accuracy depends on the order adopted \cite{Zhang:2016urt}.
Here, we avoid these possible issues as the maximum redshift is $z=0.15$~\cite{Riess:2016jrr}.
Figure~\ref{z2} shows the percentage difference with respect to the computation that considers the second-order correction, together with the distribution of the Supernovae Ia that are used to determine $H_0$. The weighted error is 0.2\%, negligible compared to the error budget to be discussed below.
Furthermore, the fact that it is safe to neglect the second-order correction implies that this analysis is valid also for spatially curved models.

%%%%%%%%%%%%%%%%%%%%%%%%%%%%%%%%%%%%%%%%%%%%%%%%%%%%%    
%%%%%%%%%%%%%%%%%%%%%%%%%%%%%%%%%%%%%%%%%%%%%%%%%%%%%

%%%%%%%%%%%%%%%%%%%%%%%%%%%%%%%%%%%%
%%%%%%%%%%%%%%%%%%%%%%%%%%%%%%%%%%%%
\section{Effective calibration prior}  \label{sec:cp}

The determination of $H_0$ by the SH0ES Collaboration is a two-step process.
On one hand the anchors, Cepheid hosts and calibrators are combined to produce a constraint on $M_B$. This step depends on the astrophysical properties of the sources and is independent of cosmology.
On the other hand the Hubble-Flow supernovae of the Supercal sample~\cite{Scolnic:2015eyc} are used to probe the luminosity distance-redshift relation in order to determine $H_0$.
In this Section, using two different methods, we will determine the calibration prior $f(M_B)$ which was effectively used in the latest 2019 analysis~\cite{Riess:2019cxk}.

\subsection{Demarginalization} \label{bayesian}

Within a Bayesian framework, the determination of $H_0$ by the SH0ES Collaboration can be effectively summarized according to the following analysis:
\begin{align}
f(H_0,M_B | \text{SN}) &= \frac{f(H_0) f(M_B) \L(\text{SN} | H_0, q_0,M_B) }{\E} \,, \label{post1} \\
f(H_0 | \text{SN}) &= \int \d M_B \, f(H_0,M_B | \text{SN})  \label{posth1} \,,
\end{align}
where the posterior on $H_0$ was obtained by marginalizing over $M_B$.
In the equations above, $f(H_0)$ is an improper flat prior on $H_0$, $\L$ is the likelihood and $\E$ is the evidence. SN stands for Supercal supernovae in the range $0.023\le z\le 0.15$~\cite{Scolnic:2015eyc}.
$f(M_B)$ is the informative prior on the supernova absolute magnitude and is the result of the complicated calibration of the local supernovae via the cosmic distance ladder, see \cite{Riess:2016jrr} for details.

The likelihood is given by:
\begin{equation} \label{newlike}
\L(\text{SN} | H_0, q_0,M_B) = |2 \pi \Sigma|^{-1/2}  e^{-\frac{1}{2}\chi^{2}(H_0, q_0,M_B)} \,,
\end{equation}
where the $\chi^2$ function is:
\begin{equation} \label{chi2sn}
\chi^{2}= \{ m_{B,i}  - m_B^t(z_i) \}  \Sigma^{-1}_{ij}  \{ m_{B,j}  - m_B^t(z_j) \} \,,
\end{equation}
where $\Sigma$ is the supernova covariance matrix and $m_{B,i}$ are the observed apparent magnitudes at the redshifts~$z_i$.
Following the methodology of \cite{Riess:2016jrr}, we fix the nuisance parameters that control stretch and color corrections to  $\alpha=0.14$, $\beta=3.1$, correct the apparent supernova magnitudes with hosts above and below $\log M_{\rm stellar} \sim 10$ by 0.03 mag fainter and brighter, respectively, and include an intrinsic dispersion of $\sigma_{\rm int}=0.1$ mag together with a peculiar velocity error of 250 km/s.

The analysis of \cite{Reid:2019tiq} fixes $q_0=-0.55$ in equation~\eqref{post1} and obtains:%
\begin{align} \label{Re19}
H^{\rm Re19}_0 = 73.5 \pm 1.4 \text{ km s}^{-1} {\rm Mpc}^{-1} \,.
\end{align}
Ref.~\cite{Riess:2019cxk} states that the sensitivity of $H_0$ to knowledge of $q_0$ is very low as the mean SN redshift is only 0.07.
However, $q_0=-0.55$ is justified only if one uses information beyond the local universe that can constrain the properties of dark matter and dark energy  (see equation~\eqref{q0}), such as data that include supernovae at $z>0.15$~\cite{Betoule:2014frx,Scolnic:2017caz}.
This was necessary as there were not enough local supernovae and/or their calibration was less precise so that it was not possible to constrain well $H_0$ if $q_0$ were not fixed. However, as we will show below~\citep[see also][]{Feeney:2017sgx}, the latest supernova catalog and anchors allow us to perform an analysis which is only based in the $z\le 0.15$ universe.
It is important to stress that the actual analysis of \cite{Riess:2019cxk} is more complicated than the one of equation \eqref{post1} as it involves many intermediate steps. 

\begin{figure}
\includegraphics[width= \columnwidth]{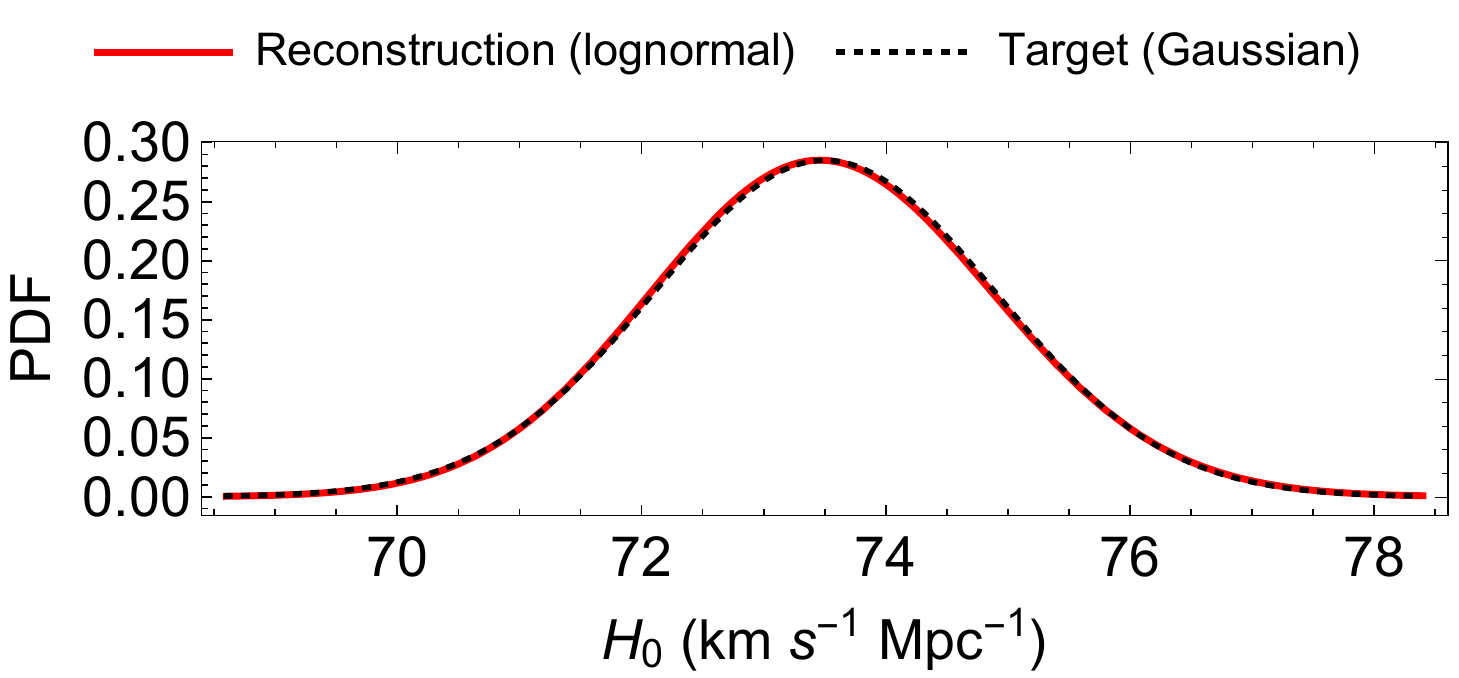}
\caption{Reconstruction of the determination by Reid {\it et al.} (2019) \cite{Reid:2019tiq} of eq.~\eqref{Re19} using the calibration prior of eq.~\eqref{prior}. Also shown is a Gaussian with same mean and dispersion. It is evident that the deviation from Gaussianity is negligible.}
\label{postH}
\end{figure}

%The determination of the informative prior on $M_B$ via the calibration of supernovae through the cosmic distance ladder \cite{Riess:2016jrr} involves a large amount of astrophysical measurements and is a non-trivial task.
Here, in order to get the calibration prior $f(M_B)$ we solve the integral equation obtained by demanding that equation \eqref{posth1} gives the constraint of equation \eqref{Re19}.
Assuming a Gaussian distribution for $M_B$ with mean $\bar M_B$ and dispersion $\sigma_M$ -- which is also justified \textit{a posteriori} by the fact that $M_B$ is tightly constrained by data -- it is possible to marginalize analytically the 2D posterior in equation \eqref{posth1}.
The result is that $H_0$ is distributed according to a lognormal distribution with parameters:
\begin{align}
\mu_{\rm ln}^{\rm dm} &= \frac{\ln 10}{5}   \left[ \bar M_B +\frac{\ln 10}{5}\left( \sigma_M^2+\frac{1}{S_0}\right)- \frac{S_1}{S_0} \right] \,,  \label{miu1} \\
\sigma_{\rm ln}^{\rm dm} &= \frac{\ln 10}{5}  \sqrt{\sigma_M^2 + \frac{1}{S_0}} \,. \label{miu2} 
\end{align}
One can then match first and second moments of the lognormal distribution with equation \eqref{Re19} so that one has:
\begin{align}
\mu_{\rm ln}^{\rm Re19} \simeq 4.2971  \,,
\qquad
\sigma_{\rm ln}^{\rm Re19}  \simeq 0.019046 \,,
\end{align}
where we used equations~(\ref{lmean}-\ref{lvar}).
The calibration prior is then given by:
\begin{align}
M_B^{\rm dm}&= \frac{5}{\ln 10} \mu_{\rm ln}^{\rm Re19}+\frac{S_1}{S_0} -\frac{\ln 10}{5}\left( \sigma_M^2+\frac{1}{S_0}\right)  \,, \label{Mb} \\
\sigma^2_{M_B^{\rm dm}} &=  \frac{25}{\ln^2 10} \sigma_{\rm ln}^{2\, {\rm Re19}} - \frac{1}{S_0} \label{sb} \,.
\end{align}
The quantities $S_0$ and $S_1$ are defined in Appendix~\ref{sec:dm}, where the full derivation can be found.
Using the previous equations one obtains the calibration prior  $f(M_B)$:
\begin{align} \label{prior}
M_B^{\rm dm} = -19.2334 \pm 0.0404 \text{ mag}  \,.
\end{align}
In the computation above we considered the second-order correction in $d_L/z$ and fixed $j_0=1$ because~\cite{Riess:2019cxk} does so, see equation~\eqref{cg3}. The second-order correction shifts the prior by a small~$\sim0.1\sigma$.
The lognormal distribution on $H_0$ is very close to a Gaussian as shown in Figure~\ref{postH}, where the lognormal $f(H_0 | \text{SN})$ is compared with a Gaussian, both with mean and dispersion as in equation~\eqref{Re19}.
%The numerical package \texttt{CalPriorSNIa} used to obtain equation~\eqref{prior} is available at \href{https://github.com/valerio-marra/CalPriorSNIa}{github.com/valerio-marra/CalPriorSNIa}.

%%%%%%%%%%%%%%%%%%%%%%%%%%%%%%%%%%%%%%%%%%%%%%%%%%%%%    
%%%%%%%%%%%%%%%%%%%%%%%%%%%%%%%%%%%%%%%%%%%%%%%%%%%%%

\subsection{Deconvolution}

One may obtain the calibration prior also via the following approach.%
\footnote{We thank the referee for suggesting this alternative method.}
The Hubble constant and the supernova absolute magnitude are connected through the following equation~\cite{Riess:2016jrr}:
\begin{align} \label{eq8R16}
\log_{10} H_0(q_0,j_0)   = \frac{M_B + 5a_B(q_0,j_0)  +25}{5} \,,
\end{align}
where the ``intercept'' $a_B$ is obtained via the Hubble-Flow supernovae:
\begin{align}
&a_B(z,m_B,q_0,j_0) =  \log_{10} [ c z f(z,q_0,j_0) ]- \frac{1}{5} m_B \,, \\
&f =1 + \frac{1-q_0}{2}z -\frac{1-q_0-3q_0^2+j_0}{6}z^2+ O(z^3) \label{cg3} \,,
\end{align}
where $z$ and $m_B$ are observed redshift and apparent magnitude of the supernovae.
In \eqref{eq8R16} the absolute magnitude $M_B$ and the intercept $a_B$ are independent as the former is obtained via the the astrophysical observations of anchors, Cepheid hosts and calibrators, while the latter via Hubble-Flow supernovae that probe the cosmological luminosity distance-redshift relation.
In particular, fixing $q_0=-0.55$ and $j_0=1$ affects $a_B$ but not $M_B$. This shows clearly that it is possible to derive the constraint on $M_B$ by ``subtracting'' (or deconvolving) the contribution of $a_B$ from the constraint on~$H_0$.

The constraint on $a_B$ using the Supercal sample~\cite{Scolnic:2015eyc} is~\cite{Riess:2016jrr}:
\begin{align}
a_B(q_0=-0.55,j_0=1)= 0.71273 \pm 0.00176 \,,
\end{align}
which can be obtained via~\cite{Schmelling:1994pz}:
\begin{align}
a_B =  \frac{\sum\limits_{i,j}(C^{-1})_{ij}\, a_{B,i}}{\sum\limits_{i,j}(C^{-1})_{ij}} \,,
\qquad
\sigma_{a_B}^2 = \frac{1}{\sum\limits_{i,j}(C^{-1})_{ij}} \,,
\end{align}
where $a_{B,i}=a_B(z_i,m_{B,i},q_0,j_0)$ are the intercept measurements with covariance matrix $C$.
After propagating the errors on $z$ (mostly peculiar velocity) on $m_B$, it is $C= \Sigma /25$, where $\Sigma$ is the covariance matrix for $m_B$.
Regarding $m_B$ and $\Sigma$ we follow the methodology of \cite{Riess:2016jrr} as described in the previous Section.

One can then proceed in two ways. The simplest approach is just to perform first-order error propagation:
\begin{align}
&M_B^{\rm e} = 5 ( \log_{10} H^{\rm Re19}_0 - a_B-5)=-19.2322 \,, \nonumber \\
&\sigma_M^{\rm e} = \sqrt{\left( \frac{5}{\ln 10}\frac{\sigma_{H^{\rm Re19}_0}}{H^{\rm Re19}_0} \right)^2 - 25 \sigma_{a_B}^2 }=0.0404 \,. \label{err1}
\end{align}
Alternatively, one can make a change of variable so that the higher moments are included.
Adopting Gaussian distributions for $M_B$ and $a_B$ it follows that $H_0$ is distributed according to the lognormal distribution of equation~\eqref{logh} with parameters:
\begin{align}
\mu_{\rm ln}^{\rm dc} &= \frac{\ln 10}{5}   (\bar M_B+ 5 a_B +25 ) \,, \\
%&=\mu_{\rm ln}^{\rm dm} -\frac{\ln^2 10}{25} \left( \sigma_M^2+\frac{1}{S_0}\right)   \,,  \\
\sigma_{\rm ln}^{\rm dc} &= \frac{\ln 10}{5}  \sqrt{\sigma_M^2 + 25 \sigma_{a_B}^2}\,, % = \sigma_{\rm ln}^{\rm dm}\,,
\end{align}
so that the calibration prior is given by:
\begin{align}
M_B^{\rm dc}&= M_B^{\rm dm} +\frac{\ln 10}{5}\left( \sigma_M^2+\frac{1}{S_0}\right)  \,, \label{Mf} \\
\sigma^2_{M_B^{\rm dc}} &= \sigma^2_{M_B^{\rm dm}} \label{sf} \,,
\end{align}
where we used the fact that:
\begin{align}
5 a_B +25 =- \frac{S_1}{S_0}  \,,
\qquad
25 \sigma_{a_B}^2 = \frac{1}{S_0} \, .
\end{align}
Using the previous equations one obtains the calibration prior  $f(M_B)$:
\begin{align}
M_B^{\rm dc} = -19.2326 \pm 0.0404 \text{ mag}  \,,
\end{align}
which confirms the validity of the first-order estimate of equation~\eqref{err1}.
We see that the demarginalization and deconvolution approaches give very similar results:
the variance of the prior is actually identical while the central value differs by a negligible amount:
\begin{equation}
M_B^{\rm dc} - M_B^{\rm dm} \simeq \frac{5}{\ln 10}\left( \frac{\sigma_{H^{\rm Re19}_0}}{H^{\rm Re19}_0} \right)^2 \simeq 8 \times 10^{-4} \,.
\end{equation}
We will adopt the demarginalization result as the analysis of the next Section generalizes its methodology.

\subsection{Robustness} \label{robust}

The computation of the calibration prior with the methods of the previous Sections is possible because  its determination is independent of the cosmological Hubble flow that is used to determine $H_0$.
In particular, $M_B$ is independent of the value adopted for the cosmographic parameters $q_0$ and $j_0$ as their effect is to alter the luminosity distance and not~$M_B$.
This would not be the case if SH0ES constrained Hubble flow and $M_B$ together as, in this case, $M_B$ would be correlated with the Hubble-flow supernovae.

Even though the methodology is well posed, there could be astrophysical or cosmological effects that could bias the determination of $M_B$.
A calibration bias could be produced by the coherent flow of the local supernovae, akin to cosmic variance~\cite{Camarena:2018nbr}.
However, both the determination of equation~\eqref{Re19} and the Supercal dataset have already been  corrected for such effects using velocity fields derived from galaxy density fields~\cite{Scolnic:2015eyc}.
Another bias could be due to correlations between supernovae and environmental properties such as star formation rate and metallicity. Ref.~\cite{Riess:2019cxk} concluded that these correlations are not significant so that they should not impact equation~\eqref{prior} either \cite[see, however, ][]{2018ApJ...854...24K,Rigault:2018ffm}. 
Therefore, we conclude that our determination of the $M_B$ prior is robust and correctly encodes the supernova calibration via the distance ladder.

\subsection{Calibration prior in cosmological analyses} \label{use}

The calibration prior on $M_B$ of equation~\eqref{prior} can be meaningfully used in cosmological analyses instead of the corresponding $H_0$ determination.
Indeed, there are 175 supernovae in common between the Supercal and Pantheon datasets in the range $0.023\le z\le 0.15$, and, in the standard analysis, these supernovae are used twice: once for the $H_0$ determination and once when constraining the cosmological parameters.
This induces a covariance between $H_0$ and the other parameters which could bias  cosmological inference.
The advantage of adopting  the calibration prior is that these 175 supernovae are used only once, avoiding any potential bias.

%%%%%%%%%%%%%%%%%%%%%%%%%%%%%%%%%%%%
%%%%%%%%%%%%%%%%%%%%%%%%%%%%%%%%%%%%
\section{Determination of local $H_0$}  \label{sec:h0}

Here, we wish to obtain a determination of $H_0$ that is based solely on the local universe ($z \le 0.15$) and that is truly independent of any cosmological assumptions besides the cosmological principle and the astrophysics of anchors, Cepheid hosts and calibrators.
%\footnote{See \cite{Valkenburg:2012td} for am observational test of the Copernican principle.}
We achieve the latter by considering %, instead of a Dirac delta prior, 
an uninformative flat prior on $q_0$ and marginalizing the posterior also over  $q_0$, that is:
\begin{align}
&f(H_0,q_0, M_B | \text{SN}) = \frac{f(H_0)f(q_0) f(M_B) \L }{\E} \,, \\
&f(H_0 | \text{SN}) = \int \d M_B \d q_0  f(H_0, q_0,M_B  | \text{SN}) \,,
\end{align}
where $f(q_0)$ is the flat ignorance prior on $q_0$.
The method of~\cite{Riess:2019cxk} is recovered if we set $f(q_0)= \delta(q_0 +0.55)$.
In this analysis we adopt the more recent Pantheon supernova sample~\cite{Scolnic:2017caz}, which features 237 supernovae in the redshift range $0.023\le z \le 0.15$, see Figure~\ref{z2}.
In this case the covariance matrix and the apparent supernova magnitudes do not depend on the nuisance parameters that control stretch and color corrections, and the correction for the apparent magnitudes with hosts above and below $\log M_{\rm stellar} \sim 10$ is already applied. Also, the covariance matrix includes all the uncertainties.
The results relative to the Supercal sample are given in Appendix~\ref{sec:sc}.

\begin{figure}
\includegraphics[width=\columnwidth]{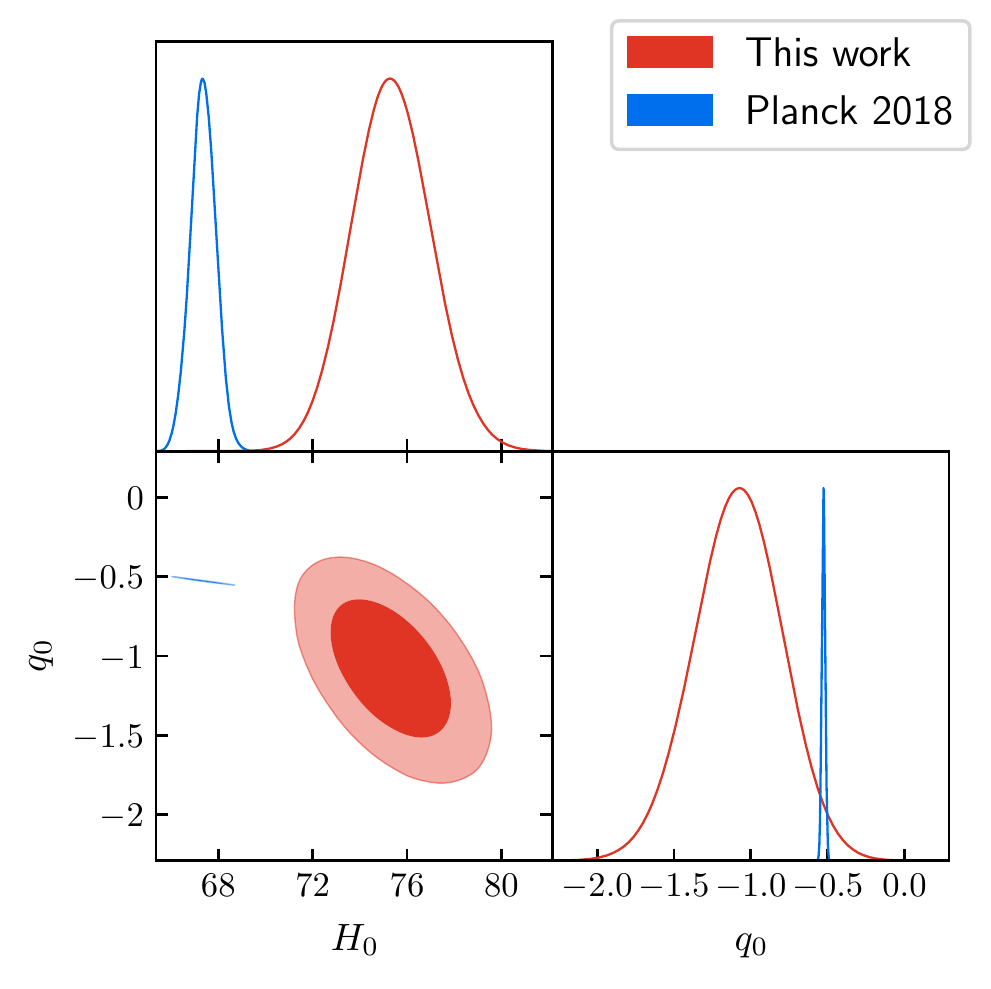}
\caption{New determination of the Hubble constant $H_0$ and the deceleration parameter $q_0$ from the Pantheon supernovae in the redshift range $0.023\le z\le 0.15$ \cite{Scolnic:2017caz}. The constraints have been marginalized over the absolute magnitude $M_B$.
Also shown are the marginalized 2D constraints on  $H_0$ and $q_0$  from the CMB-only Planck 2018 analysis that assumes the standard flat $\Lambda$CDM model \cite{Aghanim:2018eyx}. The tension  between the two determinations is at the $4.5\sigma$ level.}
\label{triplot}
\end{figure}

We have used \texttt{emcee} \cite{ForemanMackey:2012ig} -- an open-source  sampler for Markov chain Monte Carlo (MCMC) -- to sample the posterior and \texttt{getdist} \cite{Lewis:2019xzd} for analyzing the chains.
The result of this analysis is presented in Figure~\ref{triplot}. We obtain:
\begin{align}
H_0^{\rm loc} &=  75.35 \pm 1.68 \text{ km s}^{-1} {\rm Mpc}^{-1} \,, \label{H0CN} \\
q_0^{\rm loc} &= -1.08 \pm 0.29 \,. \label{q0CN}
\end{align}
The latest CMB-only constraint from the Planck Collaboration \cite[][table 2]{Aghanim:2018eyx} is:
\begin{align} \label{P18}
H^{\rm P18}_0 = 67.36 \pm 0.54 \text{ km s}^{-1} {\rm Mpc}^{-1} \,,
\end{align}
and the tension with our determination now reaches $4.5 \sigma$:
\begin{align} \label{t1}
\frac{|H_0^{\rm loc} - H^{\rm P18}_0|}{\sqrt{\sigma_{H_0}^2+\sigma^2_{H^{\rm P18}_0}}} \simeq 4.52 \,.
\end{align}
%

%%%%%%%%%%%%%%%%%%%%%%%%%%%%%%%%%%%%%%%%%%%%%%%%%%%%%    
%%%%%%%%%%%%%%%%%%%%%%%%%%%%%%%%%%%%%%%%%%%%%%%%%%%%%

As it is clear from Figure~\ref{triplot}, low-redshift supernovae are able to constrain not only $H_0$ but also $q_0$, whose distribution is peaked at values lower than the standard model one of $q_0=-0.55$.
However, the deceleration parameter is not yet tightly constrained and the tension between our determination and the one by Planck is at the $1.9\sigma$ level.
Note that the $\Lambda$CDM model cannot give values of $q_0$ below $-1$.

Finally, it is interesting to compare directly the marginalized 2D posterior on $H_0$ and $q_0$ from CMB and low-$z$ supernova observations. We obtained the CMB posterior using the MCMC chain from the CMB-only  Planck 2018 analysis available at \href{http://www.esa.int/Planck}{esa.int/Planck}.
The tension in the $H_0$--$q_0$ plane is also presented in Figure~\ref{triplot}.

In order to quantity the tension we adopt the index of inconsistency (IOI) \cite{Lin:2017ikq} which directly generalizes the estimator of equation \eqref{t1}. We find:
\begin{align}
\sqrt{2 \text{IOI}} &\equiv  \sqrt{\delta^T {(C_{\rm loc} + C_{\rm P18})}^{-1} \delta\,} \simeq 4.54 \,, \label{t2} \\
\delta &= \{H_0^{\rm loc}-H_0^{\rm P18},\;  q_0^{\rm loc} -q_0^{\rm P18} \} \,, \nonumber
\end{align}
where $C$ are the covariance matrixes on $H_0$ and $q_0$ from the analysis of Figure~\ref{triplot} and $\delta$ is the difference vector.
Note that these estimators assume Gaussianity and that the posteriors on $H_0$ and $q_0$ are very close to Gaussian. The tension in the $H_0$--$q_0$ plane is again very strong: $4.5 \sigma$.

%%%%%%%%%%%%%%%%%%%%%%%%%%%%%%%%%%%%
%%%%%%%%%%%%%%%%%%%%%%%%%%%%%%%%%%%%
\section{Discussion}  \label{sec:disc}

Our determination of the Hubble constant of equation~\eqref{H0CN} is based on the calibration prior of equation~\eqref{prior} and on the low-redshift cosmographic expansion of equation~\eqref{dL1}.
While the validity of the calibration prior depends on the standardizable nature of supernovae Ia, a process that includes corrections due to color, stretch and host-galaxy mass, the validity of the cosmographic analysis is solely based on the approximation that the FLRW metric provides a good description of our universe at large scales.
While it is possible to test directly this hypothesis~\cite{Valkenburg:2012td,Scrimgeour:2012wt,Marra:2017pst,Goncalves:2018sxa}, the validity of the FLRW metric is a direct consequence of assuming the Cosmological principle, according to which the universe is homogeneous and isotropic at large scales.
Therefore, as far as cosmological assumptions are concerned, the determination of equation~\eqref{H0CN} only assumes the Cosmological principle.

While our goal here was to improve the local determination of $H_0$ by considering a flat prior on $q_0$, it is also possible to perform a more thorough re-analysis of the cosmic distance ladder and consider the impact of non-Gaussianities~\cite{Feeney:2017sgx}, hyper-parameters~\cite{Cardona:2016ems} and a blinded pipeline~\cite{Zhang:2017aqn}.
In particular, our results are in agreement with the findings of \cite{Feeney:2017sgx}, which, using a Bayesian hierarchical model, also found that data prefer a lower value of $q_0\approx -1$ and so -- owing to their anti-correlation -- a higher value of $H_0$.
In particular, \cite{Feeney:2017sgx} adopted a broad (truncated) Gaussian prior $q_0=-0.5\pm 1$, similar to our improper flat prior, and obtained a value of $H_0$ which is $\sim1\text{ km s}^{-1} {\rm Mpc}^{-1}$ higher than their baseline $H_0$ value.
This is in good agreement with our results relative to the Supercal sample, the one adopted by \cite{Feeney:2017sgx}.
Indeed the determination of equation~\eqref{H0sc} is $\sim1\text{ km s}^{-1} {\rm Mpc}^{-1}$ than the $H^{\rm Re19}_0$ of equation~\eqref{Re19}.

It is also possible to extend the cosmic ladder to BAO and CMB observations.
Ref.~\cite{Macaulay:2018fxi} constrained $H_0$ via a cosmographic inverse ladder approach,
using BAO measurements to break the degeneracy between $M_B$ and $H_0$, and finding a  value of the Hubble constant very close to the Planck one: $H^{\rm M18}_0 = 67.8 \pm 1.3 \text{ km s}^{-1} {\rm Mpc}^{-1}$.
The cosmographic inverse ladder approach propagates the prior on the sound horizon from CMB to the BAO scale. However, the calibration of $M_B$ depends on the luminosity distance-redshift relation of supernovae both at low and high redshift.
This introduces correlations between $M_B$ and the properties of dark energy, which can explain why our cosmology-independent determination of $q_0$ differs from the constraint $q^{\rm M18}_0=-0.37 \pm 0.15$ obtained by~\cite{Macaulay:2018fxi}.

Finally, a low local value of $q_0\approx -1$ was also obtained in \cite{Camarena:2019rmj}, where the cosmic ladder was extended to BAO and CMB observations using the distance-duality relation instead of a high-redshift cosmographic expansion.

%%%%%%%%%%%%%%%%%%%%%%%%%%%%%%%%%%%%
%%%%%%%%%%%%%%%%%%%%%%%%%%%%%%%%%%%%
\section{Conclusions}  \label{sec:conc}

We obtained the effective  local calibration prior on the absolute magnitude $M_B$ of Supernovae Ia, which we used to obtain a determination of the Hubble constant from the local universe that assumes only large-scale homogeneity and isotropy: $H_0= 75.35 \pm 1.68 \text{ km s}^{-1} {\rm Mpc}^{-1}$.
Our determination uses the latest Pantheon sample.
As shown in Figure~\ref{triplot}, the allowed values are in stronger tension with what the CMB predicts if the standard model of cosmology is valid. The $H_0$ crisis is approaching the $5\sigma$ level.
We stress that our determination relies only on supernovae in the redshift range $0.023\le z\le 0.15$.
If on one hand this limits the available information as there are many more supernovae at higher redshifts, on the other hand this restricted redshift range makes this analysis local.

We would like to stress that the calibration prior on $M_B = -19.2334 \pm 0.0404$~mag can be meaningfully used in cosmological analyses, instead of the corresponding $H_0$ determination, in order to avoid the double counting of low-redshift supernovae.

The fact that  both Hubble constant and deceleration parameter differ
seems to suggest that a new phenomenon is affecting both the sound horizon and the  recent expansion rate of the universe.
Early dark energy may be a promising candidate~\cite{Poulin:2018cxd}.
The increased statistical significance of our determination of  $H_0$ reinforces this suggestion.

%%%%%%%%%%%%%%%%%%%%%%%%%%
\begin{acknowledgments}
%%%%%%%%%%%%%%%%%%%%%%%%%%

%It is a pleasure to thank XYZ for useful comments and discussions.
DC thanks CAPES for financial support.
VM thanks CNPq and FAPES for partial financial support.

%%%%%%%%%%%%%%%%%%%%%%%%%%
\end{acknowledgments}
%%%%%%%%%%%%%%%%%%%%%%%%%%

\appendix

\section{Derivation of the demarginalization}  \label{sec:dm}

Here, we explain how we obtained equations (\ref{miu1}-\ref{miu2}).
First, we define:
\begin{equation}
\tilde m_B^t \equiv m_B^t -M_B + logh \,,
\end{equation}
where we used the shorthand notation $logh =5 \log_{10} H_0 $.
We can then rewrite equation~\eqref{chi2sn} as:
\begin{align}
\chi^{2} &= \{ W_i+ logh- M_B  \}  \Sigma^{-1}_{ij}  \{ W_i +logh- M_B  \}  \nonumber \\
&= S_0(M_B - logh - S_1/S_0)^2  + const \,,
\end{align}
where $const$ consists of all the terms that do not depend on $H_0$ and $M_B$, and where we defined:
\begin{align}
W_i &=m_{B,i} - \tilde m_B^t(z_i) \,, \nonumber \\
V_i &= 1 \,, \nonumber \\
S_{0} &=  V \cdot \Sigma^{-1} \cdot V^{T} \,, \nonumber \\
S_{1} &=  W \cdot \Sigma^{-1} \cdot V^{T} \,.
\end{align}
The unnormalized posterior of equation~\eqref{post1} is then:
\begin{align} \label{postfull}
f(H_0,M_B | \text{SN}) \propto e^{-\frac{(M-\bar M_B)^2}{2 \sigma_M^2}}  e^{-\frac{1}{2}\chi^{2}(H_0,M_B)} \,,
\end{align}
where an improper flat prior on $H_0$ was adopted.
Equation~\eqref{postfull} can then be integrated over $M_B$ so that the normalized posterior on $H_0$ is distributed according to the lognormal distribution:
\begin{align}
f(H_0 | \text{SN}) &= \int \d M_B \, f(H_0,M_B | \text{SN}) \nonumber \\
&=  \frac 1 {H_0 \sqrt{2\pi}  \sigma_{\rm ln} } \exp -\frac{\left (\ln H_0-\mu_{\rm ln} \right)^2}{2\sigma_{\rm ln} ^2},  \label{logh}
\end{align}
with parameters:
\begin{align}
\mu_{\rm ln}^{\rm dm} &= \frac{\ln 10}{5}   \left[ \bar M_B +\frac{\ln 10}{5}\left( \sigma_M^2+\frac{1}{S_0}\right)- \frac{S_1}{S_0} \right] \,,  \\
\sigma_{\rm ln}^{\rm dm} &= \frac{\ln 10}{5}  \sqrt{\sigma_M^2 + \frac{1}{S_0}} \,,
\end{align}
Finally, mean and variance of the lognormal distribution are:
\begin{align}
\bar H_0  &= e^{\mu_{\rm ln} + \frac{1}{2}\sigma_{\rm ln}^2} \,, \label{lmean} \\
\sigma^2_{H_0} &=(e^{\sigma_{\rm ln}^2} - 1) e^{2\mu_{\rm ln} + \sigma_{\rm ln}^2} \label{lvar}  \,.
\end{align}
%

%%%%%%%%%%%%%%%%%%%%
\section{Analysis with the Supercal sample}  \label{sec:sc}

\begin{figure}
\includegraphics[width=\columnwidth]{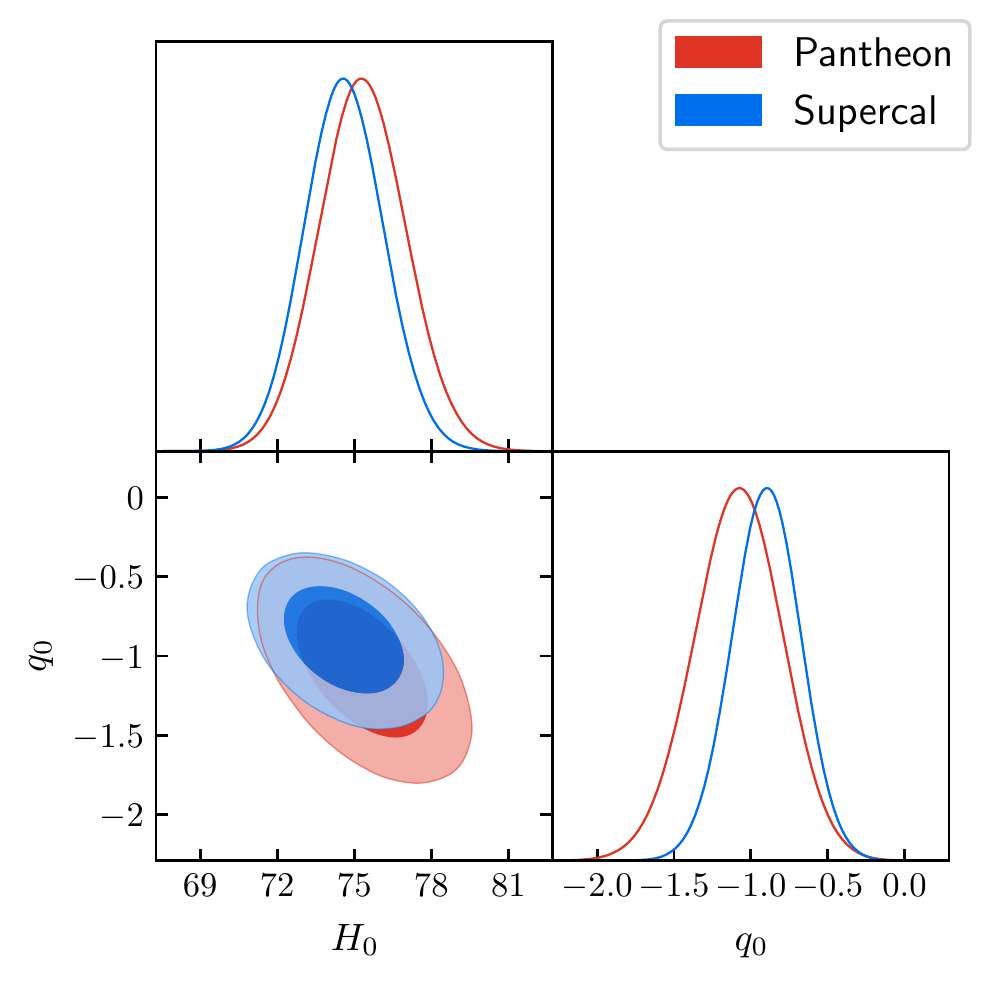}
\caption{
Comparison between the analysis that uses the Supercal supernova sample and the one of Figure~\ref{triplot} that uses the Pantheon Sample. The agreement is good.}
\label{triplot2}
\end{figure}

Here, we present the results using the Supercal supernova catalog~\cite{Scolnic:2015eyc} that was used by \cite{Riess:2016jrr,Riess:2019cxk}. It features 217 supernovae in the redshift range $0.023\le z \le 0.15$, see Figure~\ref{z2}.
Figure~\ref{triplot2} shows the results of this analysis, which are compared with the ones of Section~\ref{sec:h0} that use the more recent Pantheon sample.
We obtain:
\begin{align}
H_0^{\rm loc} &=  74.62 \pm 1.54 \text{ km s}^{-1} {\rm Mpc}^{-1} \,,  \label{H0sc} \\
q_0^{\rm loc} &= -0.90 \pm 0.22 \,,
\end{align}
which are in tension with the determinations of $H_0$ and $q_0$ by Planck at the  $4.5\sigma$ and $1.7\sigma$ levels, respectively. The tension in the $H_0$--$q_0$ plane is $4.5\sigma$.
There is good agreement between the results that use the Supercal and Pantheon datasets.

The previous results were obtained using the methodology of \cite{Riess:2016jrr}, that is, fixing the nuisance parameters that control stretch and color corrections to  $\alpha=0.14$, $\beta=3.1$, correcting the apparent supernova magnitudes with hosts above and below $\log M_{\rm stellar} \sim 10$ by 0.03 mag fainter and brighter, respectively, and including an intrinsic dispersion of $\sigma_{\rm int}=0.1$ mag together with a peculiar velocity error of 250 km/s.
If instead we adopt a flat prior on $\alpha$ and $\beta$ we obtain basically the same results:  
\begin{align}
H_0^{\rm loc} &=  74.85 \pm 1.57 \text{ km s}^{-1} {\rm Mpc}^{-1} \,,  \label{H0sc2} \\
q_0^{\rm loc} &= -0.98 \pm 0.24 \,.
\end{align}
%

%%%%%%%%%%%%%%%%%%%%%%%%%%%%%%%%%%%%%%%%%%%%%%%%%%%%%    
%%%%%%%%%%%%%%%%%%%%%%%%%%%%%%%%%%%%%%%%%%%%%%%%%%%%%

%%%%%%%%%%%%%%%%%%%%
\section{Analysis relative to Riess {\it et al.} 2019}  \label{sec:R19}

Here, we report the results relative to the early-2019 measurement by Riess {\it et al.}~(2019)~\cite{Riess:2019cxk}:
\begin{align} \label{R19}
H^{\rm R19}_0 = 74.03 \pm 1.42 \text{ km s}^{-1} {\rm Mpc}^{-1} \,.
\end{align}
The calibration prior is:
\begin{equation}
M_B^{\rm dm} = -19.2178 \pm 0.0407 \text{ mag} \,,
\end{equation}
which gives the determinations:
\begin{align}
H_0^{\rm loc} &=  75.89 \pm 1.70 \text{ km s}^{-1} {\rm Mpc}^{-1} \,, \label{H0CNr} \\
q_0^{\rm loc} &= -1.08 \pm 0.29 \,, \label{q0CNr}
\end{align}
which are in tension with the corresponding CMB-only constraints from the Planck Collaboration at the $4.8 \sigma$ and $1.9\sigma$ levels, respectively. The tension in the $H_0$--$q_0$ plane is at the $4.8 \sigma$ level.

\bibliographystyle{utphys}
\bibliography{biblio}

\end{document}